\documentclass[article,10pt,secnumarabic,showkeys]{revtex4}

\usepackage{graphicx}
\usepackage{mathrsfs}
\newcommand{\tn}{\textbf}
\newcommand{\mn}{\mathbf}

\newcommand{\rara}{\mathscr}

\usepackage{float,graphicx}
\usepackage{verbatim}

\usepackage{amsmath}
\usepackage{amssymb}

\begin{document}

\title{\small Phase ordering and symmetries of the Potts model}
\author{M. IB\'A\~NEZ DE BERGANZA$^{\S}
$\footnote{Telephone numbers and e-mails:\\ +39 0649913450,
miguel@pil.phys.uniroma1.it (corresponding author)\\ +39 0649913437, loreto@roma1.infn.it \\ +39 0649934112,
alberto.petri@isc.cnr.it}, V. LORETO$^{\S}$ and A.
PETRI$^{\ddag}$} \affiliation{$^{\S}$Dipartimento di Fisica,
Universit\`a di Roma ``La Sapienza''. Piazzale A. Moro 2, 00185
Roma, Italy.\\$^{\ddag}$CNR, Istituto dei Sistemi Complessi, sede
di Roma 2-Tor Vegata. Via del Fosso del Cavaliere 100, 00133 Roma,
Italy.}

\begin{abstract}
We have studied the ordering of the $q$-colours Potts model in two
dimensions on a square lattice. On the basis of our observations
we propose that if $q$ is large enough the system is not able to
break global and local null magnetisation symmetries at zero
temperature: when $q<4$ the system forms domains with a size
proportional to the system size while for $q>4$ it relaxes towards a
non-equilibrium phase with energy larger than the ground state energy, in
agreement with the previous findings of De Oliveira et al. \cite{PetriEL,Petri}.
\end{abstract}
 \maketitle

\section{Introduction}
\noindent
Phase ordering \cite{Lifshitz,Bray} is one of the important topics in non-equilibrium statistical mechanics. For systems with two coexisting phases the situation is generally well understood from the analytic and numeric points of view \cite{Bray,Gunton,RefsIsing}. For the Potts model \cite{Wu} with a $q>2$-degenerated ground state, the situation is not so clear in general. Also in this case, an Allen-Cahn power-law regime \cite{Grest88} and dynamical scaling relations for the structure factor and for the distribution of domain sizes
\cite{Lau,Jeppesen,Sire} have been predicted \cite{Bray} and found numerically \cite{Potts-numeric}. On the other hand, a singular behaviour of the Potts model has been recently observed by De Oliveira et. al. \cite{Petri,PetriEL} when the degeneracy ($q$) is large enough: in the thermodynamic limit the model was shown to
relax towards a `glassy', disordered, phase with a non negligible density of defects when it is quenched at zero temperature.

This paper is devoted to discuss some ideas for understanding this glassy phase. In the next section we recall the phase and report new results on the ordering of the Potts model after a quench at low but finite temperature. The system equilibrates locally nucleating domains that eventually become of the size of system. In section 3 we study the $T=0$ case, and we interpret the impossibility to equilibrate in the thermodynamic limit as an impossibility of breaking the local zero magnetisation. Section 4 is to conclude and summarise the main results.

\section{Dynamics with thermal fluctuations}
\noindent
De Oliveira et. al. \cite{PetriEL,Petri} have recently observed an
interesting slow relaxation in the dynamics of the Potts model
after a quench from a disordered state to zero temperature, and
the impossibility for the system to achieve the ground state in
the $L \to \infty$ limit, being $L$ the linear size of the system.
Let us briefly describe this effect. Given a lattice $\mn{L}$, in
which each site $i \in \mn{L}$ can take $q$ equivalent values, or
\textit{colours}, the Potts model \cite{Wu} is defined by the
Hamiltonian:

\begin{equation}
H=\frac{1}{2}\sum_{\{i,j\}}(1-\delta_{c_i,c_j})
\end{equation}
where $\{i,j\}$ means $i \in \mn L$, $j$ is a neighbour of $i$,
and where $c_i \in \{1 \cdots q\}$ is the colour of site $i$. When
quenching at zero temperature a two-dimensional system on a square
lattice using Glauber single spin-flip dynamics, systems with
$q>4$ obey the Allen-Cahn law \cite{Bray} $e(t) \propto t^{-1/2}$
($e$ being the energy per site) up to a time $\bar t$, increasing with $L$, when they get trapped in a blocked configuration, invariant with respect to
the single spin-flip dynamics at zero temperature \cite{Spirin,Godreche,Derrida96}\footnote{There exists a fraction
of realisations which fall to the ground state, but this fraction
decreases very fast with $L$ and it is negligible in the
thermodynamic limit.}. Inverting the usual order of the $t \to
\infty$ and $L \to \infty$ limits, i.e., assuming that the
thermodynamic limit is taken before the infinite-time limit, and
extrapolating to $t \to \infty$ the $e(t)$ data with the
Allen-Cahn power-law $e(t) \propto t^{-1/2}$ \cite{PetriEL}, one
finds a positive energy $e^* \equiv e(t \to \infty)>0$ for $q>4$
and, in particular the data for $e^*$ is very well fitted by the
expression $e^*=bL^{-1/2}+b'(q-4)^{1/2}$, $b,b'$ being real
constants. This would imply that an infinite-size system with $q >
4$ relaxes after an infinite time towards a phase with stationary
observables and positive energy, different from the ground state
with zero energy, and this is the reason for which this
out-of-equilibrium `phase' is called `glassy phase' in
\cite{Petri}. A noticeable point is that the onset of this
glass-like phase occurs in the two-dimensional Potts model for
$q>4$, and this $q$-range coincides with the one in which the
system presents a first-order phase transition \cite{Wu}. A
justification of this fact is in progress \cite{ILP}. We now give
a general characterisation of the equilibration of the system at finite temperature, and an interpretation of the non-equilibrium phase in terms of the symmetries of the problem in Section 3.

We have studied the dynamics after a quench at finite temperature,
$T=0.1$, of the $7$ colours Potts model with periodic boundary
conditions, being $T_c=\ln^{-1}(1+q^{1/2})=0.7730...$ the critical
temperature of the model \cite{Wu}. In Fig. 1 we present the
energy per site (on top) and the magnetisation (centre) with
respect to $t^{-1/2}$ of a 2-d square lattice system with
nearest-neighbours interaction and periodic boundary conditions.
The magnetisation, $m$, is

\begin{equation}
m=\frac{q}{q-1} \big( \sum_c^q x_c^2 -1/q \big)
\end{equation}
where $x_c$ is the fraction of colour $c$, $x_c=N_c/N$, $N_c$
being the number of sites in the system with colour $c$ and
$N=L^2$ is the total number of sites. The magnetisation is zero
when $x_c=1/q\ \forall c$, and one when all spins have the same
colour.  Initially the system is in an uncorrelated configuration
at infinite temperature, with $e=2(q-1)/q$ and $m=0$. For
$t<\tau$, systems with $q>4$ coarsen with equal proportion of all
colours and with the expected power-law dependence of the energy
on time \cite{Grest88} while, for $t>\tau$, finite-size systems
equilibrate, breaking the $m=0$ symmetry and approaching gradually
the ground state with $m=1$, $e=0$. The above description of the
problem suggests that, for $q>4$ and for times larger than $\tau$, coalescence effects, not considered in the derivation of the Allen-Cahn law, become relevant in the dynamics, and allow the mean domain size, $\ell$ \cite{Bray}, to grow up to the system size, $L$. The numerical results reported in Fig. 1 suggest that, for $T>0$, $\tau$ is not divergent in $L$, but constant above a certain $L$, i.e., that the ``nucleation'' needed to equilibrate the system is a local process no longer dependent on $L$. In fact, the $L=10^3$ and $L=500$ curves of Fig. 1 coincide within $\sigma_e^2(t,L)$, the variance of the distribution of energy values corresponding to different realisations of the quench (shown in error bars).

We observe from Fig. 1 that the $m=0$ symmetry is broken later in
larger systems. As in \cite{Fialkowsky} for the continuous Ising
model, the magnetisation is supposed to be zero for all times in
the absence of finite-size effects, and in fact we find \cite{ILP}
a similar scaling relation $m(t,L)=m(t/L^2,1)$ for $q=2,3$,
and even a slower dependence of $m(t,L)$ on $L$ for $q>4$. One
could ask why $\tau$ seems to approach a limit value with $L\to
\infty$, while the magnetisation at fixed times tends to zero in
this limit. To answer this question we propose the following
argument: systems with $q>4$ equilibrate leaving at $t=\tau$ the power-law dependence of the energy on time and forming domains that will eventually become system-sized, and $\tau$ seems to be a characteristictime of the model. It is this local nucleation the origin of the ordering, rather than the changes on the global colour fraction, which is a finite-size effect \cite{ILP}. In order to supply a quantitative support
to this argument we define an order parameter, $\gamma$,
accounting for the spatial ordering, which seems to characterise
the dynamical ordering more than the global magnetisation, $m$
\cite{magnetization}. Let us define $\gamma$ as a distance from
the $\mu(\mn r)=0$, $\forall \mn r$, situation, where $\mu(\mn r)$
is the magnetisation of a cell centred in $\mn r$ and of size $\lambda$, independent of $L$, in such a way that $\lambda/L\to 0$ in the thermodynamic limit. In particular,

\begin{equation}
\gamma\equiv\frac{1}{V} \int d\mn r\ \mu(\mn r)=\frac{q}{q-1} \frac{1}{V} \int d\mn r\ \big( \sum_c^q {\phi(\mn r)_c}^2 -1/q \big)
\rightarrow \frac{1}{L^2}\sum_{i\in \mn L}^L \mu \vert_{\mbox{\tiny
cell } i},
\end{equation}
where $V$ is the volume of the general system and $\phi_c(\mn r)\in [0:1]$ is the proportion of colour $c$ in the cell centered in $\mn r$. The expression at the right of the arrow in (3) is the definition of $\gamma$ in the
lattice $\mn L$ and  $\mbox{cell } i$ is a cell centred in the
position of site $i$ \footnote{In the numerical construction of $\gamma$, we have taken the nearest 24 neighbours of each site for the definition of the mentioned cell.}. For $L \to \infty$, $\gamma$ so defined is zero in the completely uncorrelated
configuration, and one in the ordered configuration, when all the
sites have the same colour and $m=1$. Since $\mu \ge 0$, $\gamma$ is
a distance, functional of $\mu(\mn r)$, and can be used as an
alternative order parameter accounting for the spatial ordering of
the system. We see in Fig. 1 (bottom) that also $\gamma$ presents
a power-law dependence on time. As expected by the argument
exposed above, $\gamma$ seems not to depend on $L$ for large $L$:
curves for $L=500$ and $10^3$ coincide within their standard
deviation limits. Moreover, $\gamma$ seems to leave the power-law
regime and converge to 1 at $\tau$.

\section{$T=0$ dynamics and symmetries of the glassy phase}
\noindent
At zero temperature, the situation is different: the above defined
time, $\tau$, needed to leave the Allen-Cahn law is divergent with
$L$, as described in \cite{PetriEL}, and an infinite-size system
is always supposed to follow the Allen-Cahn power-law. In larger
and larger systems, it is less and less likely to find a blocked
configuration or a path in phase space to the ground state. A
measure of this fact can be seen in the value of the energy
variance, $\sigma_e^2(t,L)$, which decreases with $L$ (see inset
of Fig. 3). This means that in a large system it is less likely
to find a deviation from the power-law regime, in which blocked
dynamics \cite{Spirin} or breaking  of the $m=0$ symmetry
\cite{Fialkowsky} are not present. This explains at a qualitative
level the fact that in large systems the magnetisation is broken
later (and hence the system equilibrates later), but it does not
address the fact that the energy of an infinite size system with
$q>4$ converges to a positive value for $t \to \infty$. In other
words, it does not explain why the term $e^*$ of the generalised
Allen-Cahn law $e = a t^{-1/2}+e^*$ is different from zero for
$q>4$. One could ask why systems with $q>4$ cannot converge to a
zero energy phase in the limit $L \to \infty$, even respecting the
$m=0$ symmetry, as the $q=2$, 3 cases do. We discuss this point in
the following.

Numerical simulations show that in the latest stage of the
coarsening at zero temperature, systems with $q=2$, 3 and with
large enough $L$ (to respect the $m=0$ symmetry) present final
configurations formed by domains of the size of system (Fig. 2,
left) and $\ell(t)$, the mean domain size, is proportional to $L$
when $t \to \infty$. The energy, or the perimeter of the
interface, is proportional to $L$, and thus, the energy per site
is zero in the thermodynamic limit. In fact we have $e^*(q)=0$ for
$q=2$, 3, as said before. On the other hand, for $q>4$, we observe
that the system presents not only $m=0$, but also a local symmetry
that is  a local equal fraction of all colours. We define this
local symmetry by introducing a certain scale, $\lambda$, with
$L>\lambda>\ell$, such that the magnetisation is also zero in
every cell of the system of size greater than $\lambda$ and \textit{being $\lambda$ such that $\lambda/L\to 0$ for $L\to\infty$}. We propose, and we argue below,  that the zero temperature dynamics cannot form
domains of the size of the system for $q>4$, hence the perimeter
of the interface (as well as the final number of domains) grows
with $L^2$ and the energy per site of the glassy phase is not zero
in the thermodynamic limit, while, in this limit, systems with
$q=2,\ 3$ do not break the $m=0$ symmetry when forming domains of
the size of the system but they can break the local $m=0$ symmetry. A numerical confirmation of this argument is that $\gamma$ converges to $\gamma^*=1$ in the $L,t \to \infty$ limits for $q=2,\ 3$, but $\gamma*<1$ for $q=7$: the fraction of the volume $V$ for which $\mu \ne 1$ is not negligible in the thermodynamic limit. In Fig. 3 we report the energy,
magnetisation and order parameter of local ordering as a function
of $t^{-1/2}$ for the $T=0$ dynamics in the $q=7$, 3 and 2 cases.
It seems that $\gamma$ presents the same power-law dependence on
time as the energy in the $q=7$ case. Inverting the order of
the limits and extrapolating to $t \to \infty$ with a linear fit,
as done for the energy in \cite{PetriEL}, we have
$\gamma^*=0.807\pm 0.005$, while for $q=2,3$, $\gamma$ converges
to 1.

This local zero magnetisation for $q>4$ (see Fig 2, right)
characterises the glassy phase also in the sense that it predicts
a $q^{1/2}$ dependence of $e^*$, which coincides in the $q \to
\infty$ limit with the result $e^* \sim (q-4)^{1/2}$ of
\cite{PetriEL}. Let us show this fact in the approximation of
domains of identical size  $\ell$ \cite{approximation}. In order
to satisfy the local-$m=0$ condition there must be $nq$ domains
inside each cell of size $\lambda$, $n$ being an integer. In $d$
dimensions it is $e\sim n_D\ \ell^{d-1}/L^d=\ell^{-1}$, where
$n_D=L^d\ell^{-d}$ is the number of domains, and setting
$\lambda^d/\ell^d=nq$ we have $e\sim \lambda^{-1} q^{1/d}$. When
$q$ approaches 4, coalescence effects not considered in this
argument become important, to which we attribute the slower dependence on $q$ of the energy, $e^* \sim (q-4)^{1/2}$.

As we have seen there are evidences that, in absence of thermal
fluctuations, the Metropolis dynamics cannot break the local
symmetry of colours when $q>4$. This fact and the $q>4$ limit can
be explained (not yet in a rigorous fashion) as a consequence of
an assumption on the glassy phase symmetries. To introduce it let
us define the $q$ `coarse-grained' fields $\phi_c(\mn r) \in
[0,1]$, in analogy with \cite{Sire}, as the fraction of colour
$c$ in a cell centred in $\mn r$ and of size $\lambda$. Clearly
$\sum_c^q \phi_c=1$ and the fields $\phi_c$ describe the
configuration in a $\lambda$-dependent way. Given the fact that
the Hamiltonian is invariant with respect to $S_q$, the group of
permutations of the $q$ colours, we assume that the $m=0$
symmetric configuration corresponding to $L\to\infty$, $t \to
\infty$ is maximally ordered in such a way that any permutation of
the colours, $p\in S_q$, is equivalent to a $p$-dependent global
spatial transformation, $U_p$. i.e.:

\begin{eqnarray}
\phi_{p(c)}(\mn r)=\phi_c(U_p\mn r) \hspace{0.5cm} \forall p \in
S_q
\end{eqnarray}
If we assume the operator $U$ to be linear we have that $U: S_q
\to \rara{L}(\mathbb R^2)$ is a representation of $S_q$ in the
vector space of linear operators in the plane, $\rara{L}(\mathbb
R^2)$. Since there is no representation of $S_q$ in
$\rara{L}(\mathbb R^2)$ for $q>4$ different from the trivial
representation ($U_p=\mathbb I$, the identity operator on $\mathbb
R^2$, $\forall p$), from the above assumption it follows the
local-zero magnetisation of the glassy phase for $q>4$: setting
$U_p=\mathbb I$ in (4) we obtain $\phi_{c'}(\mn r)=\phi_c(\mn r)$,
that with the normalisation condition gives $\phi_{c}(\mn r)=1/q$
$\forall \mn r,c$, i.e., the local $m=0$ symmetry in the cell of
size $\lambda$. The assumption (4) can also `justify' the spatial
symmetries of configurations with $q \le 4$ in the $m=0$ regime:
for $q=2$ the group $S_2$ admits a nontrivial representation in
the plane, $\{\mathbb I,\alpha\}$, where $\mathbb I$ is the
identity and $\alpha$ is the rotation of $\pi$ radians in the
polar angle and, in fact, configurations with $q=2$ in the glassy
phase seem to present this symmetry (see Fig. 2, left), and a
similar $2\pi/3$ rotational symmetry seem to exist for $q=3$
configurations, even if it is often hindered by the two-preferred
directions of the square lattice interaction \cite{Spirin} and by
the formation of structures locally stable under the $T=0$
dynamics \cite{SafranI,Lifshitz}. Moreover, in a general
$d$-dimensional system, there exist no nontrivial representations
of $S_q$ in $\mathbb R^d$ for $q \ge d+2$  \cite{groups}. This proposition seems
to be true for all $q>4$, and we are working on the general proof.
If verified, this consequence of the assumption (4) would coincide
with the result by Lifshitz \cite{Lifshitz}, who argued that a
$d$-dimensional system quenched below its critical point does not
necessarily equilibrate into an ordered state, in the presence of
a ground state that is degenerated more than $d+1$ times.

\section{Conclusions and further research}
\noindent 
We have studied the ordering dynamics of the 2-d Potts
model in a square lattice with $q=2,3$ and 7 by use of single-spin
flip dynamic Monte-Carlo simulations. At positive temperature,
$T=0.1$, and for $q=7$ the system relaxes leaving the Allen-Cahn power law at a given time, which is independent on $L$ for large $L$. At zero temperature, in the $L \to \infty$ limit, systems with $q>4$ are not able to nucleate breaking
the local symmetry, and hence they converge to a phase with nonzero
energy. We give a quantitative support for this argument in the
$q=7$ case with the help of an order parameter of local ordering
that we define. The $q>4$ limit is presented as a consequence of a
hypothesis on the symmetries of the glassy phase. For both
temperatures the magnetisation at equal times decreases with
increasing sizes and it is argued to be zero in the thermodynamic
limit at any time \cite{Fialkowsky}. A generalisation of this study for different
$q$ values and for larger $L$ is being performed \cite{ILP}
together with an investigation of the relation between the
ordering properties and the phase space structure.\\

\begin{figure}[h]
\begin{center}
\includegraphics[height=8.5cm]{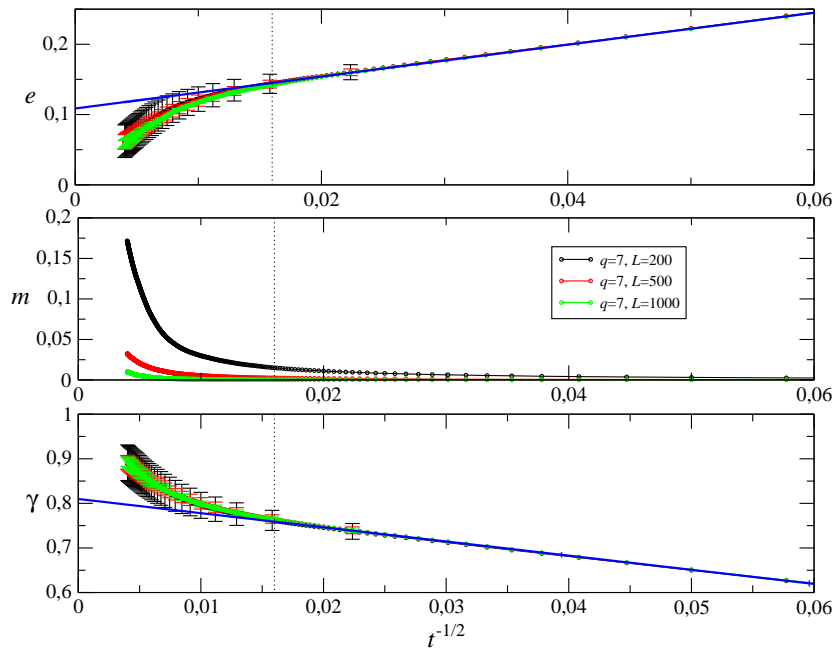} \\
\end{center}
\caption{\footnotesize Energy per site ($e$), magnetisation ($m$)
and order parameter of local ordering ($\gamma$) with respect to
$t^{-1/2}$ for the Potts model with $q=7$ after a quench to
temperature $T=0.1$. Results are averaged over 200, 60 and 30
realisations (for $L=200,\ 500$ and $10^3$, respectively). Time is
in MCS units. Error bars in the $y$-axes are the variances, $\sigma^2$, of $e$ and $\gamma$, corresponding to the
average over different realisations of the quench (shown every
2000 MCS only). In larger systems the $m=0$ symmetry is broken
later \cite{Fialkowsky}. $\sigma^2$ decreases with the size of the
system at equal times. The blue line is a fit to the $L=10^3$
system energy in the range $t^{-1/2}\in [0.02,0.1]$, and the
vertical dotted line is an estimation of the nucleation point,
corresponding to the time at which $e$ differs from the fit more than
$\sigma_e^2/2$. At that time, $\tau$, the system nucleates leaving
the Allen-Cahn power law, and $\gamma$ increases towards 1.}
\end{figure}

\begin{figure}[h]
\begin{center}
\begin{tabular}{c c}
\includegraphics[height=3.5cm]{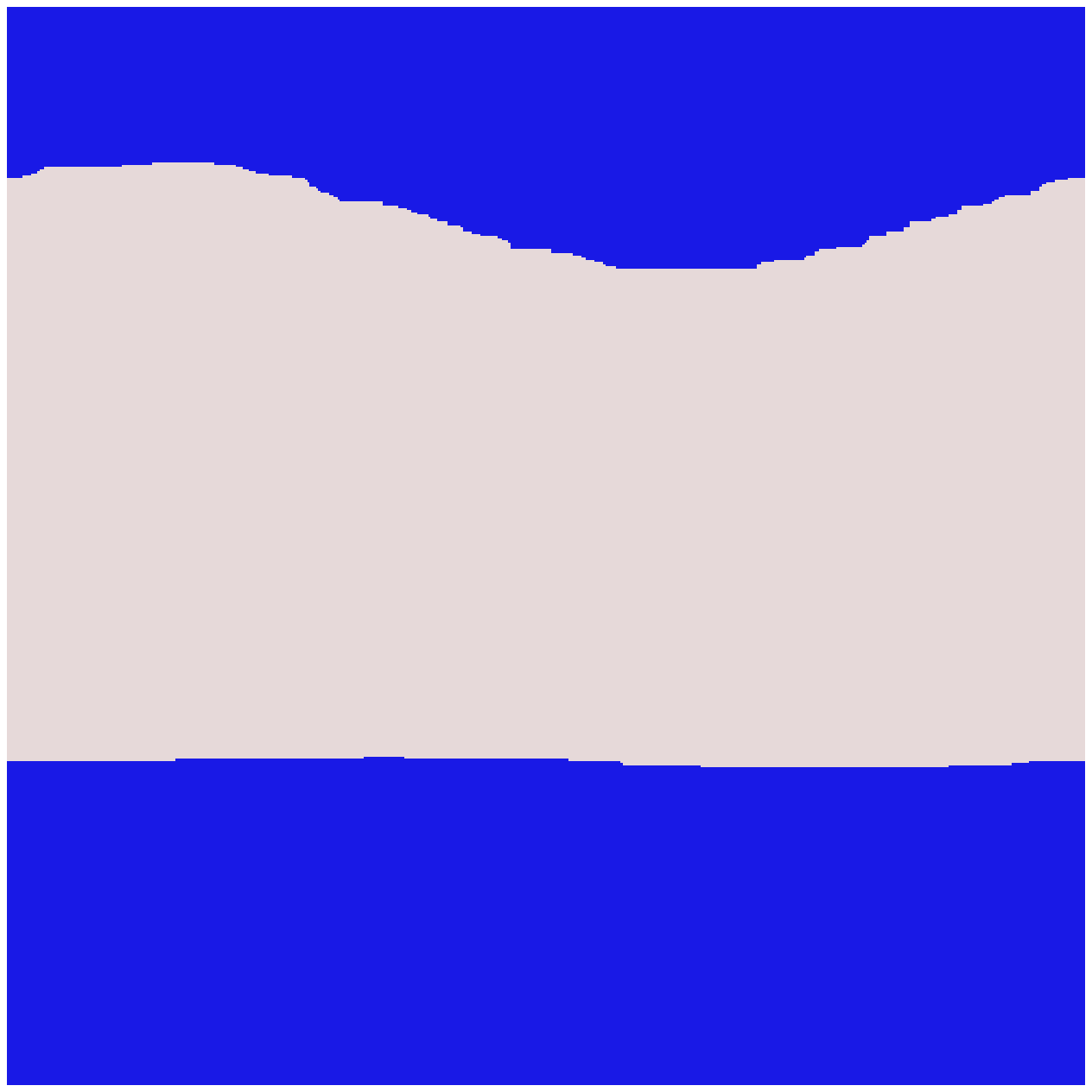} &
\includegraphics[height=3.5cm]{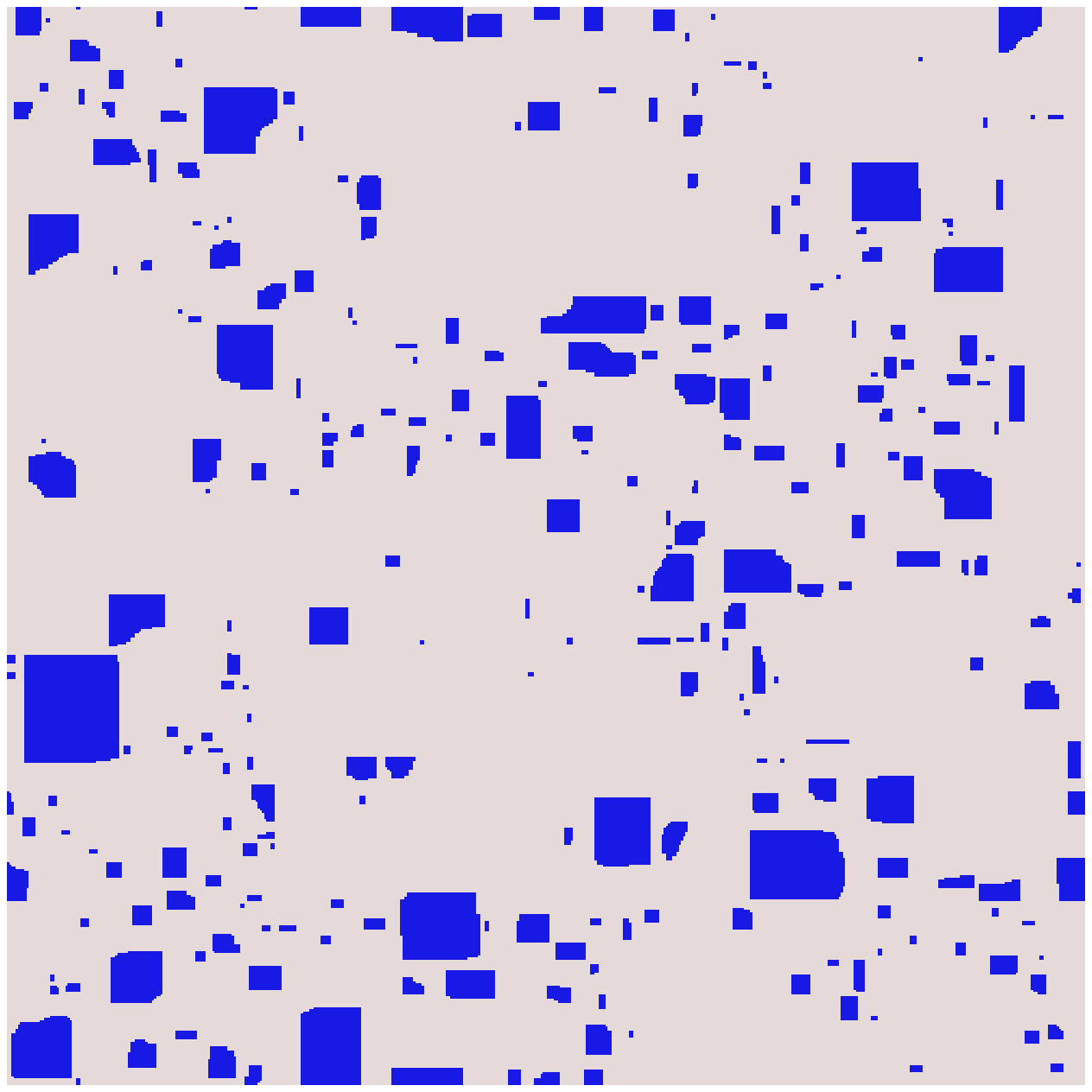} \\
$q=2$ & $q=7$ \\
$\gamma=0.992$ & $\gamma=0.775<\gamma^*$
\end{tabular}
\end{center}
\caption{\footnotesize Example of two configurations reached by the 2 and 7-colours Potts in a $L=500$ lattice, after $6.25\ 10^4$ MCS of a single realisation of a quench at zero temperature. The magnetisation is $m=1.9\ 10^{-4}$ and $9.6\ 10^{-4}$, respectively. Spins with a given colour are shown in blue, while spins with the remaining $q-1$ colours are shown in white. In the latest stage of the coarsening the interfaces between domains of the $q=2$ case become straight lines.}
\end{figure}

\begin{figure}[h]
\begin{center}
\includegraphics[height=8.5cm]{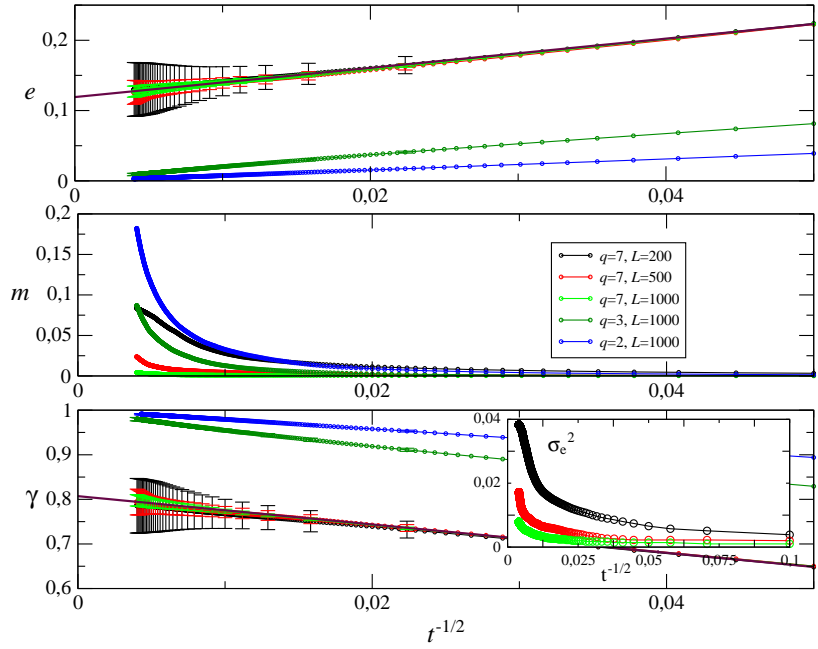} \\
\end{center}
\caption{\footnotesize Energy per site ($e$), magnetisation ($m$) and order parameter of local ordering ($\gamma$) with respect to $t^{-1/2}$ for the Potts model after a quench at zero temperature. The $q=7$, 3 and 2 cases are presented, and three different sizes for the $q=7$ case. Magnitudes are an average over 240, 60 and 40 realisations of the quench in systems with $L=200$, 500 and $10^3$, respectively.  For the cases $q=2$ and 3 the energy converges to zero following the Allen-Cahn law, while in the $q=7$ case it converges to $e^*= 0.1192$ (\cite{PetriEL}). The global magnetisation decreases with $q$ at equal times and sizes, for all sizes and times studied. It also decreases with $L$ for all times \cite{Fialkowsky}, and it is supposed to be zero in the thermodynamic limit. Even with this constraint, systems with $q<4$ can form system-sized domains with zero energy, and in fact they order spatially, and $\gamma(t \to \infty)\to 1$, while the zero-temperature dynamics cannot break the local zero magnetisation symmetry to form domains of the size of the system when $q>4$, as illustrated by the fact that $\gamma(t \to \infty)\to \gamma^*<1$ for $q=7$, and this is the reason for which $e^*>0$. The extrapolated line is a fit to the $L=10^3$ data for $0.005<t^{-1/2}<0.06$ MCS$^{-1/2}$. The fit coincides with the data within its variance limits. The equation of the fit is $\gamma_{\mbox{\tiny{fit}}}=\gamma^*+at^{-1/2}$ with $\gamma^*=0.8072$ and $a=-3.1756$. On inset we present the variance $\sigma_e^2$ of the energy, which is a decreasing quantity with $L$, for all times and for the three sizes studied.}
\end{figure}

\begin{thebibliography}{99}
\bibitem{approximation} The validity of the mentioned approximation is discussed in \cite{ILP}.
\end{thebibliography}
\end{document}